# Religious Festivals and Influenza

Alice P.Y. Chiu[1], Qianying Lin[1], Daihai He[1,*]

[1]Department of Applied Mathematics, Hong Kong Polytechnic University,

Hong Kong (SAR) China

*Correspondence to: daihai.he@polyu.edu.hk

TU804, Yip Kit Chuen Building, Department of Applied Mathematics,

Hong Kong Polytechnic University, Hong Kong (SAR) China

# Abstract

## Objectives

Influenza outbreaks have been widely studied. However, the patterns between influenza and religious festivals remained unexplored. This study examined the patterns of influenza and Hanukkah in Israel, and that of influenza and Hajj in Bahrain, Egypt, Iraq, Jordan, Oman and Qatar.

## Method

Influenza surveillance data of these seven countries from 2009 to 2017 were downloaded from the FluNet of the World Health Organization. Secondary data were collected for the countries' population, and the dates of Hajj and Hanukkah. We aggregated the weekly influenza A and B laboratory confirmations for each country over the study period. Weekly influenza A patterns and religious festival dates were further explored across the study period.

## Results

We found that influenza A peaks closely followed Hanukkah in Israel in six out of seven years from 2010 to 2017. Aggregated influenza A peaks of the other six Middle East countries also occurred right after Hajj every year during the study period.

## Conclusions

We predict that unless there is an emergence of new influenza strain, such influenza patterns are likely to persist in future years. Our results suggested that the optimal timing of mass influenza vaccination should take into considerations of the dates of these religious festivals.

**Introduction**

Globally, influenza accounts for a substantial mortality and morbidity annually. Influenza patterns have been extensively studied at regional and global scale. He et al. (2015) explored the global spatio-temporal patterns of influenza dynamics[1]. Axelsen et al. (2014) studied the seasonal influenza dynamics in Israel, and found that they were driven by humidity, temperature, immunity lost and antigenic drift[2]. Huppert et al. (2012) examined the spatio-temporal patterns of seasonal influenza in Israel and found that both the timing and the strength of the influenza outbreaks were highly synchronized between the cities[3]. Barnea et al. (2014) analysed the spatio-temporal synchrony of influenza in Israel and concluded that all cities generally followed the same epidemic curve each season[4]. Some of these studies examined influenza-like-illness (ILI) data[2,3,4], while ILI data did not necessarily well match the influenza laboratory-confirmed data. Both ILI and laboratory-confirmed data could have pros and cons in describing the true epidemic.

In Israel, Hanukkah is a Jewish holiday that involves mass gathering. In the Middle East region, Hajj attracted millions of pilgrims from all over the world to gather at Mecca in Saudi Arabia[5]. Our aim is to study the impacts of these religious festivals on the patterns of influenza. Specifically, we explored the patterns of influenza during Hanukkah in Israel, and that of Hajj in six Middle East countries surrounding Saudi Arabia, i.e. Bahrain, Egypt, Iraq, Jordan, Oman and Qatar.

**Data Collection and Methods**

Influenza surveillance data were downloaded from FluNet of the World Health Organization[6] for the following countries: Bahrain, Egypt, Iraq, Jordan, Oman, Qatar and Israel. The time period covered was from January 1st, 2009 to August 20th, 2017. The number of influenza confirmations tested by type was downloaded on a weekly basis.

Population statistics of these seven countries as of 2016 were extracted from the World Bank[7]. We aggregated the influenza confirmations across the time period by influenza type and country. Temporal patterns of influenza A during religious festivals were explored. We compared the influenza A patterns of Israel versus the other six countries.

**Results**

Figure 1 summarises the population as of 2016 and the total influenza A and B laboratory specimens collected of the seven Middle East countries from Jan 2010 to Aug 2017 (exclude 2009 pandemic). The overall reported number of influenza A specimens were approximately five-fold that of influenza B.

Figure 2 shows the weekly influenza A confirmations of Israel (red curve) versus the six Middle East countries (shaded regions) from Jan 2009 to Aug 2017. The patterns are striking: in all study years but 2009, influenza A peaked after Hanukkah in Israel. Furthermore, the aggregated influenza A confirmations of the other six Middle East countries peaked after Hajj consistently each year.

## Discussion

To the best of our knowledge, our study is novel in identifying consistent influenza A peaks that occurred after religious festivals. Mass gathering during these festivals potentially contributed to influenza transmission. This explained the influenza peaks after Hajj and Hanukkah that occurred consistently during the study period. We predict unless an emergence of a new influenza strain (e.g. in 2009), such patterns between influenza and religious festivals are likely to persist in future years.

Our study has several key strengths. We used laboratory confirmed influenza surveillance data from FluNet that covered a sufficiently long time period. Also, an attempt is made to predict the likely influenza patterns in the upcoming years. However, our study is limited by the non-availability of influenza surveillance data from Saudi Arabia.

Our results have public health implications. By identifying that influenza A peaks are likely to occur after these religious festivals, the optimal timing of mass influenza vaccination should consider these festival dates. Further studies could explore the exact mechanism causing these influenza patterns.

## Figure Captions

Figure 1. A map of the Middle East region showing the populations of each country as of 2016 and their cumulative influenza A and B specimens collected from Jan 2010 to Aug 2017. Note that influenza surveillance data from Saudi Arabia were not publicly available. Mecca is where the mass gathering of Hajj pilgrimage is held annually. (The country borders are from Sandvik B., World Borders Dataset http://thematicmapping.org/downloads/.)

Figure 2. The weekly influenza A confirmations of Israel (red curve) and the aggregate of six other Middle East countries (shaded regions) from 2009 to 2017. The start dates of Hajj and Hanukkah in each year are marked. Hajj is ahead of Hanukkah, so is the arrival of influenza wave in the six countries than in Israel between 2010 and 2017.

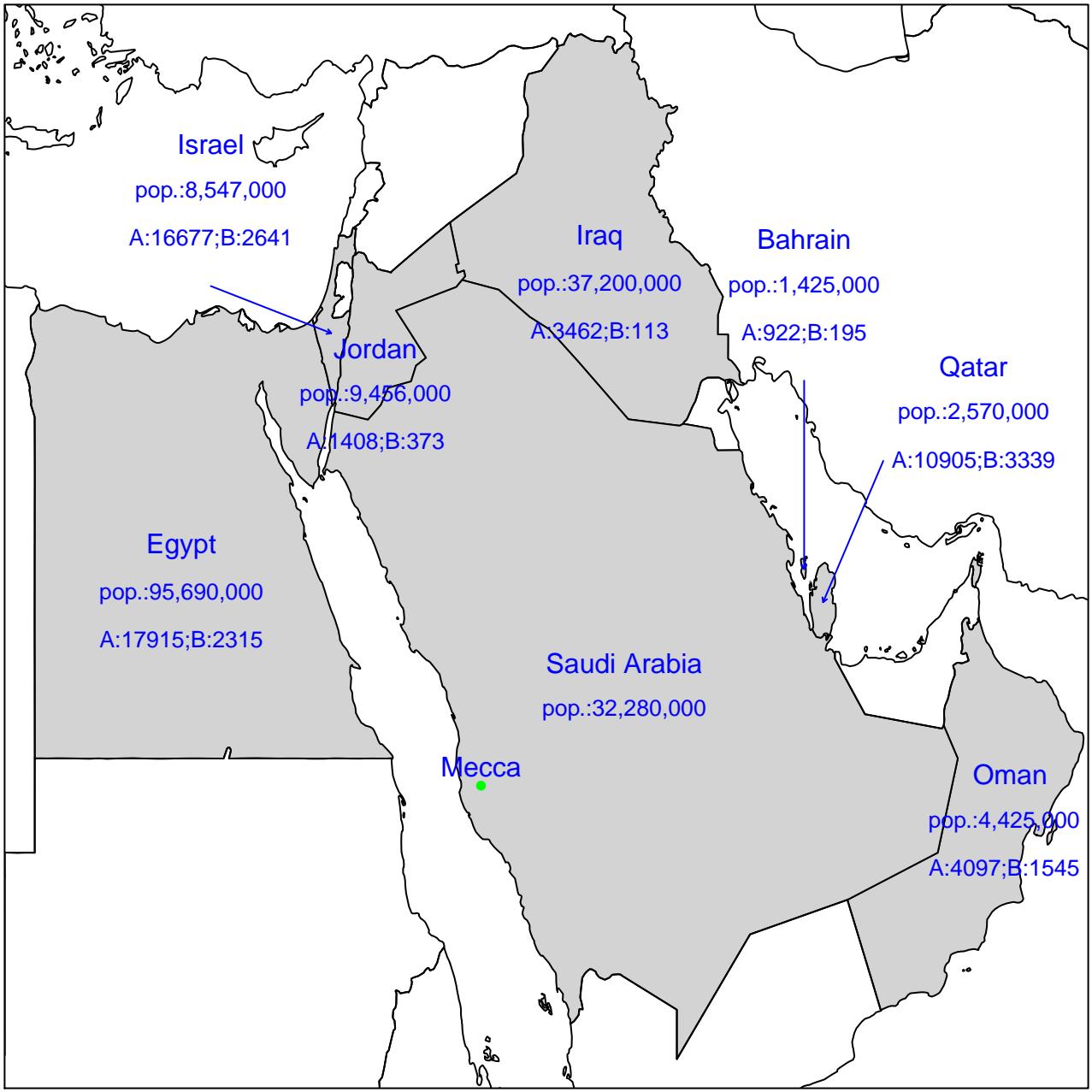

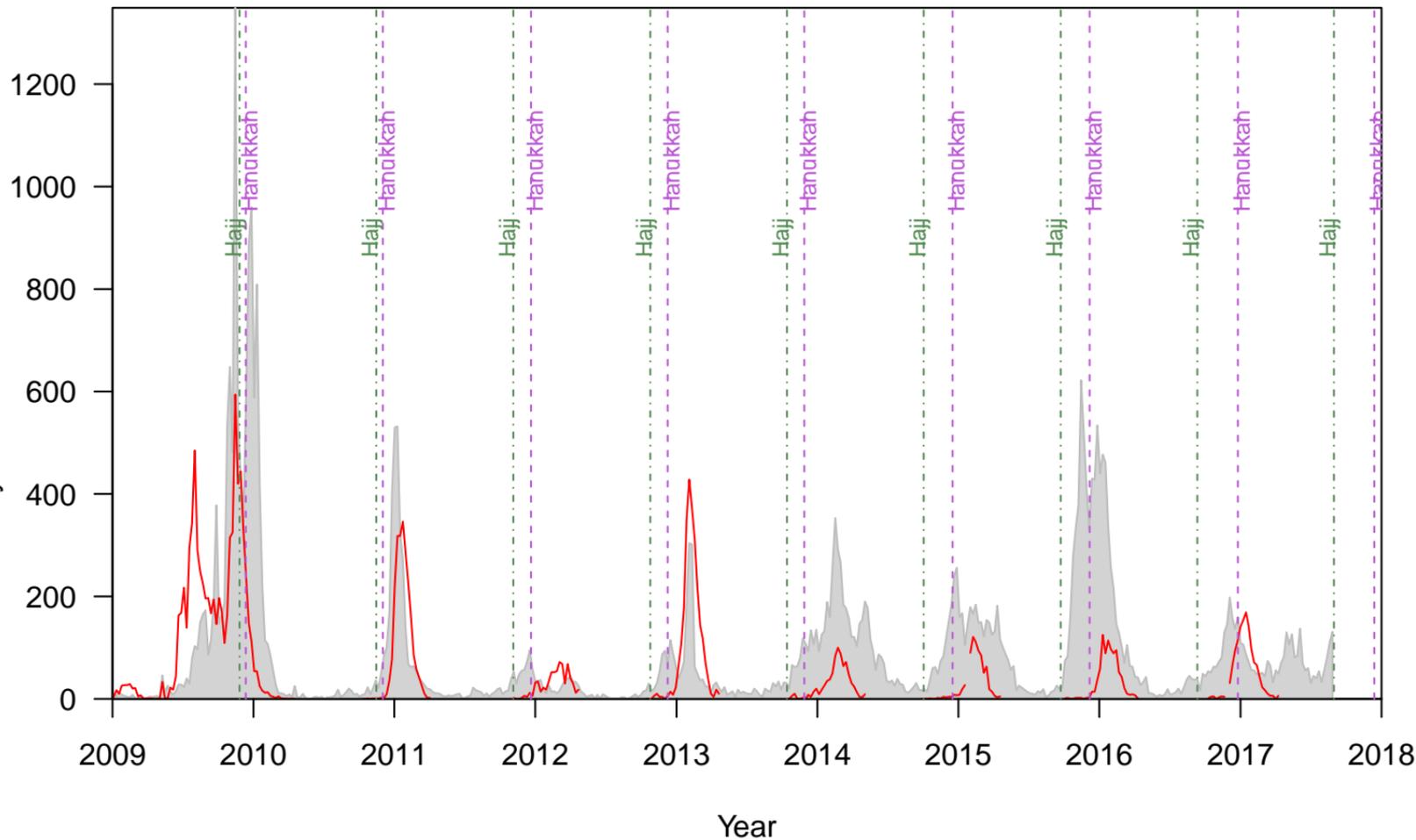